\documentclass{wiley-article}
\usepackage[super]{natbib}
\usepackage{wrapfig}
\usepackage{siunitx}
\papertype{Perspective}
\title{Molecular simulation approaches to probing the effects of mechanical forces in the actin cytoskeleton}

\author[1]{Fatemah Mukadum}
\author[2]{Willmor J. Pe\~na Ccoa}
\author[1,2]{Glen M. Hocky}

\affil[1]{Department of Chemistry, New York University, New York, NY 10003, USA}
\affil[2]{Simons Center for Computational Physical Chemistry, New York, NY 10003, USA}
\corraddress{Glen M. Hocky, Department of Chemistry, New York University, New York, NY, 10003, USA}
\corremail{hockyg@nyu.edu}
\fundinginfo{This work was supported by the National Institutes of
Health through the award R35GM138312.}

\runningauthor{Mukadum et al.}
\begin{document}
\maketitle
\begin{frontmatter}
\begin{abstract}
In this article we give our perspective on the successes and promise of various molecular and coarse-grained simulation approaches to probing the effect of mechanical forces in the actin cytoskeleton.
\end{abstract}
\end{frontmatter}

\section{Brief overview of forces in the actin cytoskeleton}
\begin{figure*}[t!]
    \centering
   \includegraphics[width=0.95\columnwidth]{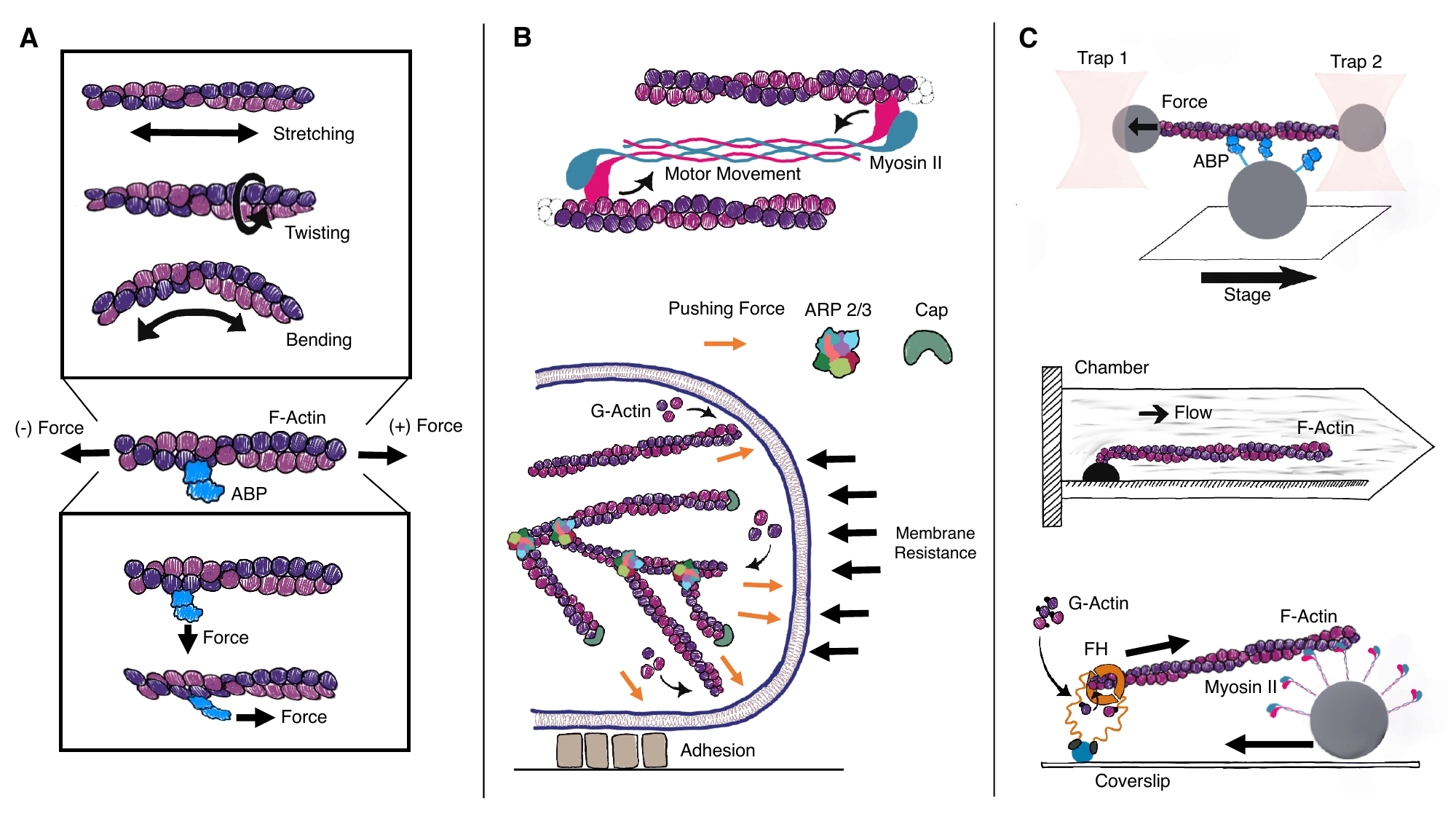}
    \caption{(A) Actin filaments are polar helical non-covalent polymers. 
    Forces applied directly to filaments can cause stretching, twisting, and bending, while forces applied to actin binding proteins (ABPs) perpendicular or parallel can cause stronger or weaker binding. 
    (B) In cells, forces on proteins are caused by motion of motor proteins along filaments, by polymerization of filaments against structures, or through interactions with cellular adhesion proteins. 
    (C) In vitro, the effect of forces on actin and actin binding proteins can be probed by using tweezers,\cite{huang_vinculin_2017} microfluidic devices,\cite{wioland_advantages_2020}  or with motors anchored either to substrates or beads.\cite{zimmermann_mechanoregulated_2017}}
    \label{fig:force-overview}
\end{figure*}

The actin cytoskeleton is a network of crosslinked polymers in cells that is a major locus of force generation and transmission.\cite{pollard_actin_2009} 
Its primary component is the filamentous actin (F-actin) polymer (Fig.~\ref{fig:force-overview}A), working in concert with numerous actin binding proteins (ABPs)\cite{dominguez_actin-binding_2004,pollard_actin_2016} that play roles in crosslinking, polymerizing, severing/depolymerizing, or nucleating filaments.\cite{pollard_actin_2016}
Forces on the cytoskeleton can be generated through the action of myosin molecular motors, which are either anchored in place and pull unidirectionally on a filament (as is the case with myosin I)\cite{laakso_myosin_2008} or are combined into a bundle and pull on multiple filaments simultaneously (as in the case of myosin II) \cite{pollard_actin_2009,svitkina_actin_2018} (Fig.~\ref{fig:force-overview}B). 
Actin itself can serve as the origin of force, as the action of polymerizing globular actin (G-actin) from solution can be used to generate pushing forces\cite{kovar_insertional_2004,marcy_forces_2004}; this is particularly relevant at the leading edge of cells, where branched actin networks nucleated by Arp2/3 complex are used to propel the cellular plasma membrane forwards (Fig.~\ref{fig:force-overview}B).\cite{blanchoin_actin_2014,liu_membrane-induced_2008} 
Generation of forces either through polymerization or motor activity require the conversion of chemical energy to mechanical energy via the hydrolysis and release of ATP molecules. 
Actin itself in solution contains an ATP molecule, and catalyzes the conversion of this ATP to ADP after undergoing a transition from a twisted to a flat state upon entering into a filament, meaning that actin filament polymerization alone can produce mechanical work.\cite{dominguez_actin_2011}

Mechanical forces experienced by actin or actin binding proteins would typically be in the range of zero to several piconewtons (pN).\cite{blanchoin_actin_2014} 
Individual myosin heads are individually capable of producing several piconewtons of force, \cite{finer_single_1994,laakso_myosin_2008,stam_isoforms_2015} meaning that in certain cases, filaments could experience hundreds of piconewtons of tension.
Pulling or twisting forces on the pN scale can be translated into down stream signaling effects, meaning it is certainly relevant to be able to predict the molecular affect of these forces on filament structure.\cite{de_la_cruz_cofilin_2005,jegou_mechanically_2021,cossio_catching_2022,reynolds_bending_2022,bibeau_twist_2023}
The effect of forces on the behavior of actin binding proteins in some cases can be probed by \textit{in vitro} experiments, including those employing molecular tweezers or microfluidic devices to controllably manipulate individual filaments in solution \cite{finer_single_1994,huang_vinculin_2017,owen_c-terminal_2022,jegou_mechanically_2021} (Fig.~\ref{fig:force-overview}C). 
However, the impact of these forces on the atomic level cannot be directly observed, and hence here we give our perspective on the use of molecular dynamics simulations to predict the molecular response of small mechanical forces on actin and actin binding proteins.

\section{Including Force in Molecular Dynamics Simulations}
Molecular dynamics (MD) simulations are a set of techniques where we seek to explicitly model the motion of molecules by treating all the atoms as if they follow the rules of classical mechanics.\cite{tuckerman_statistical_2010}
The intra and intermolecular forces dictating the motion of particles comes from a model for the potential energy of the arrangement of atoms termed a ``forcefield.'' \cite{schlick_biomolecular_2021}
External mechanical forces can be included through modification of the system's ``Hamiltonian'' (the sum of potential and kinetic energy), as
\begin{equation}
    H(\vec{x},\vec{p})=\sum_{i=1}^N \frac{\vec{p}_i^2}{2 m_i}+U_\mathrm{forcefield}(\vec{x})+U_\mathrm{mechanical}(\vec{x}),
\end{equation}
where $\vec{x}$ denotes the positions all atoms in the system, and $\vec{p}$ their momenta.

The contribution of mechanical forces to the term $U_\mathrm{mechanical}(\vec{x})$ correspond to doing mechanical work, such that for a static pulling force,  $U_\mathrm{mechanical}(\vec{x})=-F_\mathrm{mechanical}d(\vec{x})$, where $d(\vec{x})$ is the coordinate to which pulling is applied, e.g. the distance between two atoms or the distance between the centers of mass of two residues in a protein.\cite{bustamante_mechanical_2004,gomez_molecular_2021}
Experimentally, single-molecule pulling forces can be applied by attaching the molecule of interest to an AFM or an  optical or magnetic trap \cite{bustamante_single-molecule_2020}; rather than a constant force, this is typically modeled in simulation by introducing a harmonic restraint $U_\mathrm{mechanical}(\vec{x})=\frac{1}{2} k_\mathrm{trap} (d(\vec{x})-d_0)^2$ centered at a distance $d_0$.\cite{lu_steered_1999} 
In `steered molecular dynamics' (SMD), $d_0$ is often moved linearly in time to mimic constant velocity experiments.\cite{lu_steered_1999,stirnemann_recent_2022}
While in this article, we are concerned with modeling the effect of true mechanical forces, we also note that these same techniques are often applied to abstract coordinates in order to produce a first guess of a transition trajectory, e.g. to study the activation pathway of Arp2/3 complex or the phosphate release pathway from within filamentous actin.\cite{ding_structure_2022, singh_molecular_2023, dalhaimer_molecular_2010,wriggers_investigating_1999,okazaki_phosphate_2013,wang_mechanism_2023}
For now, we and others have treated forces in one of these two modalities, but one lingering question is how best we should apply forces to filaments or binding proteins to  mimic how the forces are applied in experiment or in vivo; i.e. we feel it is important to determine whether our current treatments are sufficient to capture the effects produced by shear forces or stochastic motor generated-forces.

Typically, the goal of MD simulations is not to study the true dynamics of the system but rather to `sample' configurations from the true equilibrium ensemble; to do so, the equations of motion of the system are modified to include extra terms that keep the system at either constant temperature or at constant temperature and pressure.\cite{tuckerman_statistical_2010}
When this is done, then the configurations seen in the MD simulation should arise with probability proportional to the proper statistical distribution, e.g. in the case of constant temperature, $P(\vec{x},\vec{p})\propto e^{-H(\vec{x},\vec{p})/(k_\mathrm{B}T)}$, where $k_\mathrm{B}$ is the Boltzmann constant, and $\mathrm{k_\mathrm{B}T}\approx0.593$ kcal/mol $\approx 2.479 $kJ/mol $\approx 4.114$ pN nm at room temperature (298 K).
When constant forces are applied, MD simulations performed at constant temperature or at constant temperature and pressure will still sample from this distribution.
After doing so, we often want to compute a `free energy surface' (FES) also known as a `potential of mean force' (PMF) which allows us to visualize the relative free energy of configurations along a small number of `collective variables' (CVs) represented by $\vec{Q}(\vec{x})$.
This FES corresponds to the negative log of the frequency with which configurations have a particular value $\tilde{Q}$,
\begin{equation}
    F(\tilde{Q})=-k_\mathrm{B} T \ln\left(\int \delta(\vec{Q}(\vec{x})-\tilde{Q}) P(\vec{x},\vec{p})  d\vec{x} d\vec{p} \right) + \mathrm{const.},
\end{equation}
where $\delta$ is the Dirac delta function.\cite{tuckerman_statistical_2010}

A significant challenge associated with MD simulations is that a small integration time step must be used to propagate the equations of motion, typically 2 femtoseconds for atomistic MD of proteins.\cite{tuckerman_statistical_2010,henin_enhanced_2022}
This limits the amount of sampling to times corresponding to microseconds at most, which is not enough time to sample all relevant configurations from $P(\vec{x},\vec{p})$, since relevant biological conformational transitions in which we are interested can have characteristic time scales corresponding to milliseconds or longer.
Enhanced sampling methods can help bridge this gap and allow us to compute PMFs and even the rates of slow events within available computational time, as described next.\cite{tuckerman_statistical_2010,henin_enhanced_2022}

\section{Enhanced sampling for determining the effect of forces}
In most cases, the use of enhanced sampling techniques will be indispensable for assessing the affect of force on actin and actin binding protein structure. 
A vast array of enhanced sampling techniques exist based on different approaches for quickly crossing free energy barriers.\cite{henin_enhanced_2022}
We develop and employ these approaches using the PLUMED open-source plugin to many popular MD codes, which allows us to simultaneously apply constant or time dependent forces to our system.\cite{bonomi_promoting_2019} 

For large protein complexes, we favor CV-based approaches in which coordinates for biasing are carefully chosen to characterize the states of interest, and ideally the transition state between those states. 
The original such approach is termed umbrella sampling, wherein the system is constrained (typically by a harmonic potential) to be close to a certain value of a CV; by combining information from many simulations scanned across the physically relevant range of CV values, a PMF can be reconstructed.\cite{tuckerman_statistical_2010,henin_enhanced_2022}
\begin{figure}[ht]
    \centering
    \includegraphics[scale=0.5]{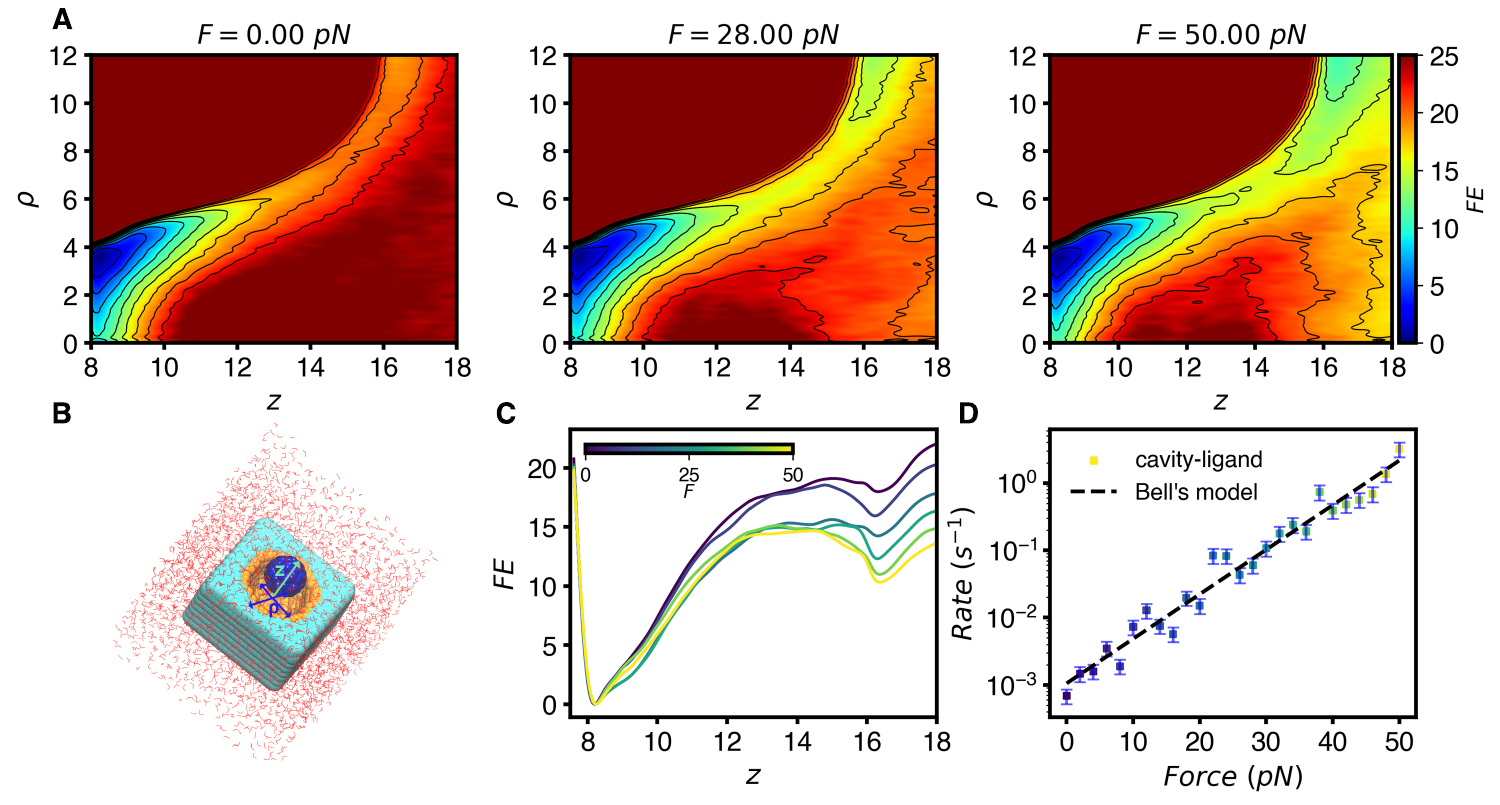}
    \caption{Example of enhanced sampling for a system with applied force, adapted from Ref.~\citenum{pena_ccoa_assessing_2022}. (A) Two-dimensional FESs computed from MetaD for a model ligand-receptor system in B. Here, the FES was computed by biasing perpendicular ($z$) and parallel motion $\rho$ to the cavity surface. Application of force to $z$ shifts the equilibrium towards unbound. (C) One-dimensional PMF after integrating out $\rho$ shows how application of force `tilts' the free energy landscape. (D) Rates computed by infrequent MetaD and fit to Bell's law.}
    \label{fig:sampling}
\end{figure}

In our work, we typically employ Metadynamics (MetaD) and similar approaches, which have emerged as a very popular method for simultaneously exploring and computing an unknown FES.\cite{bussi_using_2020,invernizzi_unified_2020}
In MetaD, a history-dependent bias potential is added to the system's Hamiltonian. 
This bias potential is formed from a sum of Gaussian `hills' that are periodically deposited at the system's current position in CV space. 
As a result, the system is driven away from previously explored regions; additionally, the amount of bias applied at each position is then used to estimate the underlying free energy surface, as the amount of bias used is proportional to the negative of the underlying FES.\cite{bussi_using_2020}
While this allows one to compute free energies much more readily than unbiased sampling, the quality of the result and speed of convergence still depends strongly on the choice of CVs.\cite{bussi_using_2020, henin_enhanced_2022}
Other variants of MetaD exist that provide distinct advantages; for example the MetaD flavor of OPES (on the fly probability enhanced sampling) progressively updates an estimate of the whole FES rather than building it from a sum of Gaussians, which can give more robust convergence.\cite{invernizzi_unified_2020} 
OPES also permits use of an energy cutoff above which bias is not applied, which can help prevent exploration of unphysical regions of phase space as was shown to be important in the sampling of actin flattening with metabasin MetaD.\cite{dama_exploring_2015}
For constant forces, MetaD can be directly applied to determine how the conformational ensemble of a system changes in response to force.\cite{gomez_molecular_2021,pena_ccoa_assessing_2022}

MetaD and related methods can also be used to compute the rates of certain very slow processes.\cite{tiwary_metadynamics_2013,ray_kinetics_2023}
If bias is added only in the starting basin and not on the transition state, then it can be shown that the effect is to accelerate time by an amount related to the exponential of the bias applied, averaged over the starting basin.\cite{tiwary_metadynamics_2013,ray_kinetics_2023}
Many such simulations can be performed and the rate can be computed from the mean of the rescaled times \cite{tiwary_metadynamics_2013,ray_kinetics_2023} or more recently through a maximum likelihood approach.\cite{palacio-rodriguez_transition_2022}
The exponential factor means that rates for processes even on the hours time scale can be obtained from simulations that are only tens of nanoseconds in length, if good CVs are chosen for biasing. 
We demonstrated that this approach is able to predict the force-dependence of unbinding rates for several different systems (see Fig.~\ref{fig:sampling}); however, difficulty arose when we tested the approach on larger protein systems due to the presence of intermediates along the unbinding pathway.\cite{pena_ccoa_assessing_2022}
The approach of Ref.~\citenum{palacio-rodriguez_transition_2022} can alleviate some of the error that can arise from choice of CVs, but getting good convergence for large biomolecular complexes such as in the actin cytoskeleton remains a challenge.

In addition to MetaD, it is also possible to develop methods more directly tailored to the effect of force on biomolecular configurations. 
We previously developed an approach termed Infinite Switch Simulated Tempering in Force, \cite{hartmann_infinite_2020} which allows us to assess the effect of a range of forces from a single simulation; we recently demonstrated that convergence of this method can be improved by combination with methods that accelerate sampling through running at multiple temperatures in parallel. \cite{singh_improved_2023} 

For larger biomolecular complexes, these kinds of CV based approaches may be insufficient, at least with currently available computational resources. 
For probing the pathway between two known states of a system, path-based methods such as the string method with swarm of trajectories \cite{e_string_2002,pan_finding_2008} have been successfully used to make reasonable predictions of minimum free-energy pathways; we believe these approaches could also be employed with and without constant force to determine how force changes the free energy barrier for transition in a much higher dimensional space. 
For larger systems, coarse-grained (CG) approaches that reduce the dimensionality of the system may be required; \cite{marrink_martini_2007,noid_perspective_2013,schlick_biomolecular_2021,jin_bottom-up_2022} while these models sacrifice some chemical detail and perhaps some of the physics contained in atomisticly detailed forcefields, it is the only reasonable approach when tackling biophysical problems at scales beyond thousands of amino-acids (or nucleotides). 
We believe such models can be leveraged to find useful trends even if they are not quantitative.

\section{Computational insights into actin mechanosensitivity}
\begin{figure*}[t!]
    \centering
   \includegraphics[width=1.0\columnwidth]{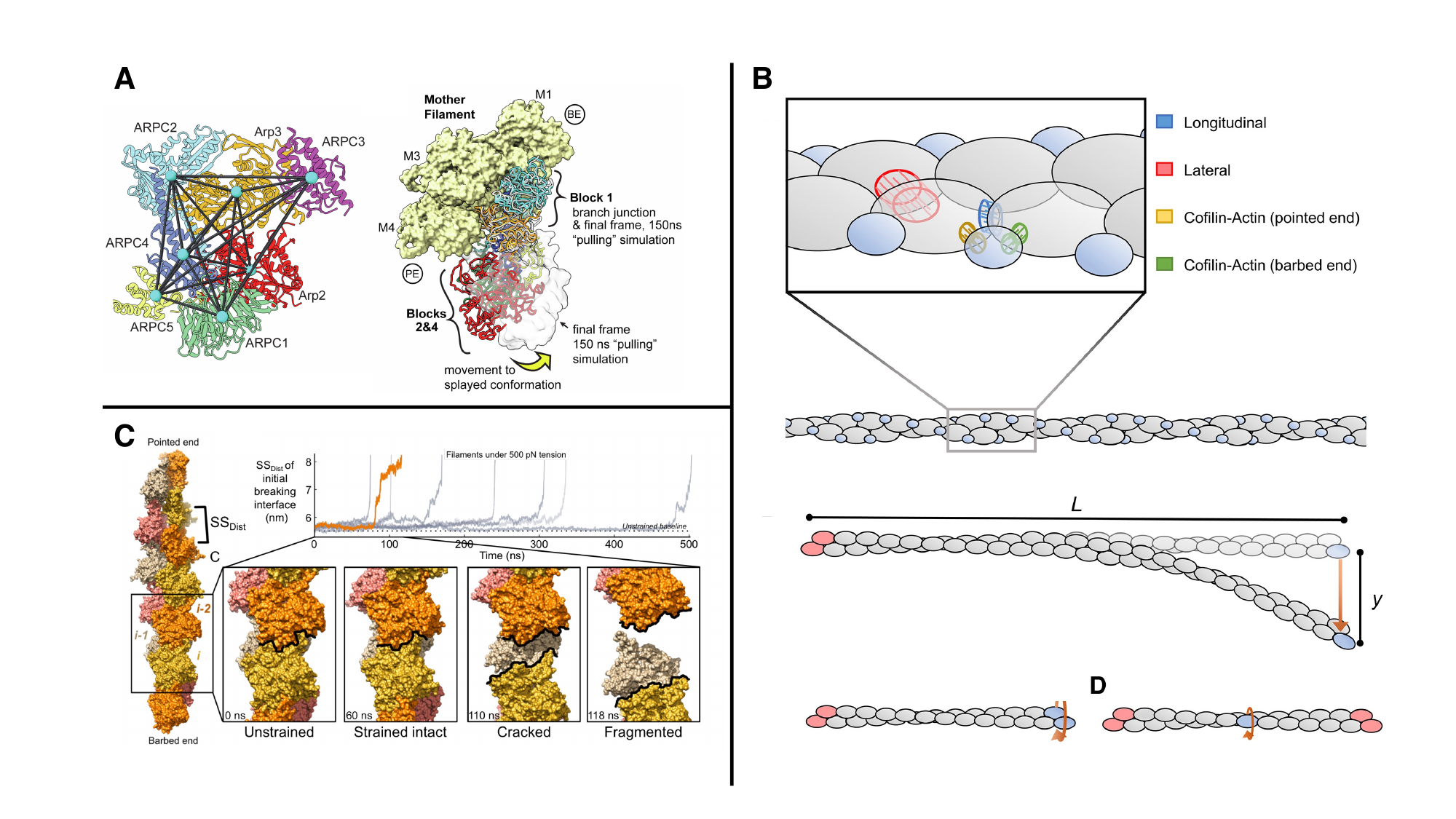}
    \caption{(A) Steered MD applied to pairwise distances between domains of Arp2/3 complex promotes transitions between active and inactive states, adapted from Ref.~\citenum{singh_molecular_2023}. 
    (B) A CG model for probing the effect of bending and twisting on cofilin-bound actin filaments, adapted from Ref.~\citenum{schramm_plastic_2019}. 
    (C) Forces applied to filaments produced cracks that could be loci for proteins targeted to regions of force, adapted from Ref.~\citenum{zsolnay_cracked_2023}.}
    \label{fig:lit-examples}
\end{figure*}

Here we give a few highlights of how MD simulations coupled to forces have already been used to give insight into actin and actin binding proteins.
As previously mentioned, forces can be introduced artificially into MD simulations in order to drive rare transitions, for example promoting unbinding events, as in the release of inorganic phosphate from actin filaments.\cite{wriggers_investigating_1999, okazaki_phosphate_2013, wang_mechanism_2023}
Forces have also been used to promote transitions of assemblies of actin and actin binding proteins, such as converting the actin branching Arp2/3 complex from from an inactive to an active state  (Fig.~\ref{fig:lit-examples}A), \cite{ding_structure_2022, singh_molecular_2023, dalhaimer_molecular_2010} or transitions between bound states of tropomyosin on actin.\cite{williams_mechanism_2018}

Application of (perhaps unrealistically large) forces have been used as a proxy for predicting the mechanical properties of actin monomers, \cite{mehrafrooz_mechanical_2018}  filaments, \cite{shamloo_nanomechanics_2018} or filaments bound to binding proteins such as cofilin.\cite{kim_steered_2016}
As an alternative to atomistic simulations, coarse grained simulations\cite{kim_steered_2016, chu_coarse-grained_2006} and mechanical models\cite{li_atomistic_2018,schramm_actin_2017,schramm_plastic_2019} have also been employed to study the bending, stretching and twisting of F-actin (e.g. Fig.~\ref{fig:lit-examples}B). 

Only recently have computational approaches been powerful enough to predict the response of atomistic models of actin to more realistic forces. 
For example, all-atom MD simulations under 500 pN of extensional strain showed metastable cracked conformations that could relate to the origin of force-activated binding of regulatory factors (e.g. Fig.~\ref{fig:lit-examples}C).\cite{zsolnay_cracked_2023}
We expect many more such studies in the future, and we point to possible areas of opportunity below in Sec.~\ref{sec:future}.

\section{Case study: F-Actin/Vinculin Complex}
We argue that we are now poised to leverage enhanced sampling techniques in the presence of small forces to probe the molecular mechanisms underlying mechanosensing behavior in the actin cytoskeleton. 
Here we give an example of how we are combining several techniques as well as large scale computational resources to tackle a problem of particular interest, namely the origin of catch bonding behavior in the focal adhesion protein vinculin.

\begin{wrapfigure}{r}{7.5cm}
\begin{center}
\vspace{-0.9cm}
\includegraphics[width=7.5cm]{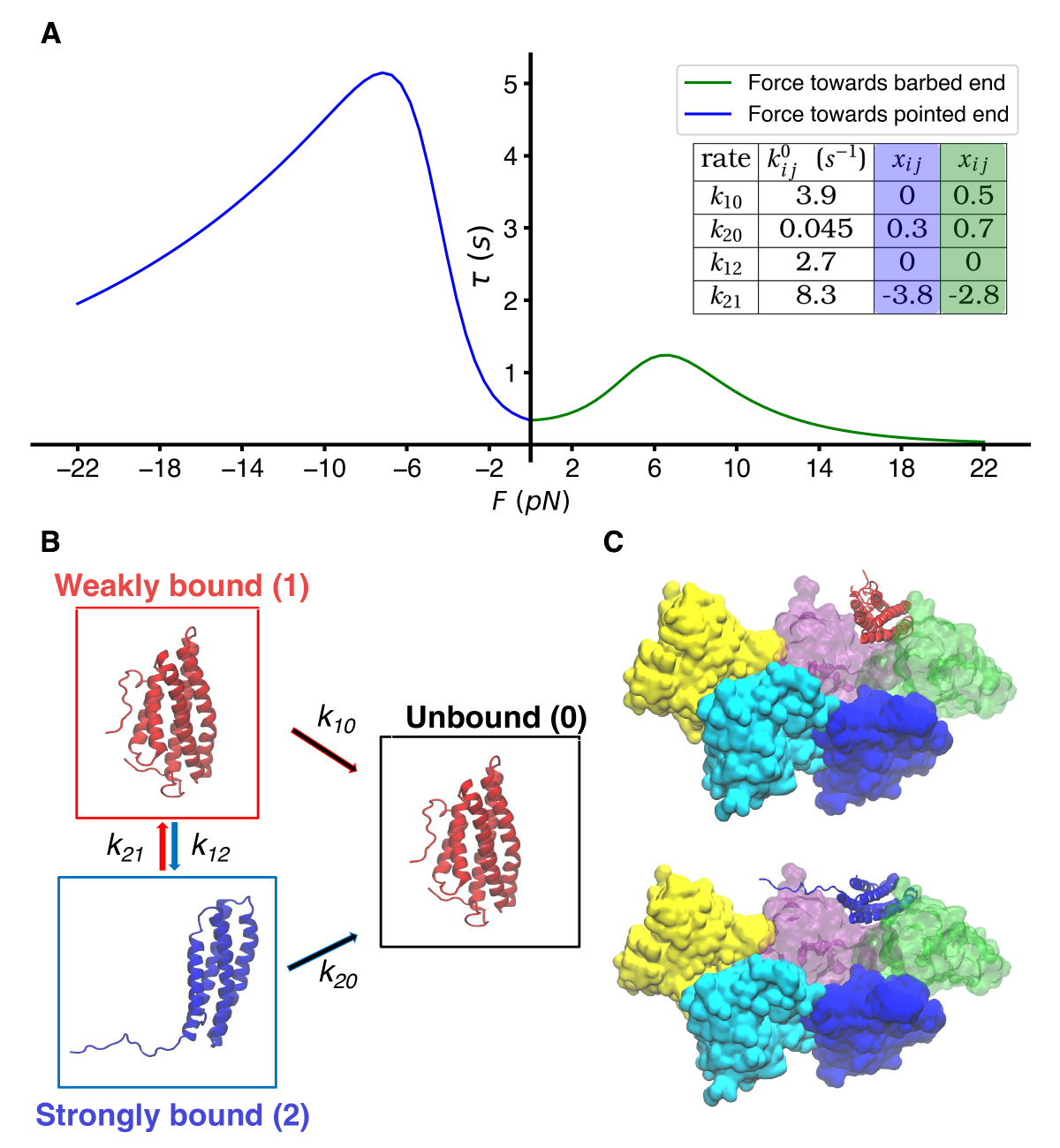}
\vspace{-0.8cm}
\caption{(A) Best-fit curves for the lifetime of Vt bound to actin as assessed in Ref.~\citenum{huang_vinculin_2017} give rise to a parameterized model (see table) with two bound states and one unbound state. (B) Vinculin four-or five helical bundles corresponding to the putative strong and weak bound states of catenin from Refs.~\citenum{wang_mechanism_2022} and \citenum{mei_molecular_2020}. (C) Some of our models of five actin subunits bound to either four and five-helical bundle states of Vt, for probing the force sensitivity of Vt unbinding.}
\label{fig:vinculin}
\end{center}
\vspace{-1cm}
\end{wrapfigure}

When force is applied across a protein-protein interaction, the naive expectation is that it will accelerate the rate of unbinding, or equivalently, decrease the lifetime of the `bond'. 
Simple theoretical arguments show that small forces should accelerate processes exponentially with force, something known as Bell's law. \cite{gomez_molecular_2021,bell_models_1978}
In contrast to this `slip bond' behavior, some biological assemblies show `catch' bond behavior, where the lifetime actually gets longer for small forces \cite{gomez_molecular_2021,chakrabarti_phenomenological_2017}. 
In focal adhesion and adherens junction assemblies, which help anchor cells to substrates, it is proposed that catch bond behavior engaged by forces applied between the cytoskeleton and the extracellular environment play a role in stabilizing attachment.\cite{huang_vinculin_2017}
Single molecule experiments have shown that proteins vinculin and catenin form a catch bond when bound to F-Actin \cite{huang_vinculin_2017,wang_mechanism_2022}.
For vinculin, catch bonding behavior was shown to exist whether the force was applied towards the pointed or barbed end of F-actin, however, the lifetimes of the interactions are larger when pulling towards the pointed end (see Fig.~\ref{fig:vinculin}A).\cite{huang_vinculin_2017}
Furthermore, it was shown that the vinculin tail (Vt, 5-helix bundle) is a key structure for the catch bond behavior, as the directional catch bond was still observed in optical trap experiments where only Vt was bound to F-Actin.

\setlength{\columnsep}{6pt}%
For both catenin and vinculin, a two-state catch bond model \cite{chakrabarti_phenomenological_2017} was used to describe the force dependence of the unbinding, consisting of an unbound state, a weakly bound state, and a strongly bound state.\cite{huang_vinculin_2017}
Within this model, each individual transition can be treated as following Bell's law, $k_{ij}(F) = k_{ij}^0 e^{F x_{ij}/(k_BT)}$,
where $k_{ij}^0$ is the transition rate from state $i$ to state $j$ in the absence of force and $x_{ij}$ is a parameter for the distance to the transition barrier between states $i$ and $j$. 
Thus the apparent catch bond behavior arises as a consequence of the rapid force-dependent transition to the strongly bound state.
Molecular models proposed for $\alpha$-catenin point to a transition between a four-helix bundle and a five-helix bundle as the key structural motifs underlying this multistate model (see Fig.~\ref{fig:vinculin}B).\cite{mei_molecular_2020,wang_mechanism_2022}
So far, little has been done from a modeling standpoint to probe the molecular underpinnings of this process or to test the hypothesis proposed through structural biological techniques. 
Very recently, MD simulations with large forces on Vt bound to two actin subunits were used to point to key interactions involved in directional bond strengthening, and when mutated these produce Vt variants that are able to bind but showed evidence of loss of catch-binding ability.\cite{chirasani_elucidation_2023}

We see this vinculin/actin interplay as an extremely challenging, yet ideal platform to test the enhanced sampling MD simulation approaches above.
In order to properly tackle all aspects of the problem, we should be able to find the force-dependent rates of unbinding, as well as the free energy barriers dictating changes between states of the protein in its bound pose with actin. 
Are simulation approaches robust enough to actually predict free energies and rates for this very slow (seconds-time scale) process?
Achieving this result requires high quality and large simulations of cytoskeletal assemblies in each of the different putative states, some of which are shown in Fig.~\ref{fig:vinculin}C.  
We are leveraging free energy and rates approaches as described above \cite{invernizzi_unified_2020,palacio-rodriguez_transition_2022}, as well as methods for characterizing relevant states of the system  and designing high quality CVs requiring minimal bias for producing transitions \cite{klem_size-and-shape_2022,sasmal_reaction_2023} aiming to develop a workflow that is able to provide robust predictions.
We are optimistic that such approaches will give detailed insight into how such catch bonding behavior arises in an atomic model of a cytoskeletal system for the first time.

\section{Future opportunities}
\label{sec:future}
Advancing in vitro and structural biological techniques are constantly improving our understanding of the actin cytoskeleton, especially in how actin and several ABPs coordinate in a dynamic fashion to self-organize.\cite{kadzik_f-actin_2020}
For example, single-particle cryo-electron microscopy approaches combined with machine learning approaches are giving new insight into how coupled bending and twisting alter the structure of individual filaments \cite{reynolds_bending_2022}, and MD simulations may be able to leverage this information to predict how these changes alter binding affinity for ABPs.
We previously showed that the cost of bending actin filaments can lead to sorting of actin crosslinkers in bundles,\cite{freedman_mechanical_2019} but these arguments were based on the persistence length of actin filaments, derived from equilibrium bending fluctuations; the ability of filaments to permit bends over shorter length scales is something that could be further detailed through computational modeling.

As another example, EM studies and MD simulations together have given complementary insights into the interface between cofilin-bound and bare actin filaments,\cite{huehn_structures_2020, hocky_structural_2021} and ability to accurately model application of force could give a detailed picture of how this leads to filament severing and depolymerization. 
Tackling larger scale force-induced processes such as the debranching of Arp2/3 complex \cite{pandit_force_2020} likely will remain out of reach by atomistic approaches for some time, but could perhaps be investigated through CG modeling.

As described, in vitro experiments show that  mechanical forces can produce changes of binding affinity for ABPs.\cite{jegou_mechanically_2021}
Recent MD studies point to a mechanism for force to expose residues within filaments, which could be a mechanism used by LIM domain proteins to sense stressed filaments, which can also play an important feedback role in maintenance of mechanical properties. \cite{zsolnay_cracked_2023,winkelman_evolutionarily_2020,anderson_lim_2021,sun_mechanosensing_2020,sun_cellular_2023}
A major challenge remaining is to demonstrate that a combination of accelerated MD simulations with small forces can have predictive accuracy for such subtle yet important structural changes.

\section*{Acknowledgments}
GMH would like to thank all past and present collaborators on these topics for many fruitful conversations, in particular Guillaume Stirnemann with whom we are collaborating on the vinculin/actin project described above. 
 FM and GMH were supported by the National Institutes of Health (NIH) via Award No. R35GM138312, and W.J.P.C. by R35GM138312-S1.

\printendnotes

\bibliographystyle{unsrtnat}
\bibliography{CytoskeletonPerspective}

\end{document}